\documentclass[journal,10pt]{IEEEtran}
\usepackage{amsmath,amssymb,verbatim,cite,amsopn,graphicx,url,color,balance}
\usepackage{algorithm}
\usepackage{algorithmic}
\usepackage{multicol}
\usepackage{epsfig}
\usepackage{epstopdf}
\usepackage{subcaption}
\usepackage{stfloats}
\usepackage[font={small}]{caption}

\makeatletter
\def\maketag@@@#1{\hbox{\m@th\normalfont\normalsize#1}}
\makeatother

\def\tablename{Table}
\renewcommand{\thetable}{\arabic{table}}

\newcommand{\hab}{h_{ab}}
\newcommand{\hba}{h_{ba}}
\newcommand{\hbi}{h_{bi}}

\newcommand{\hai}{h_{ai}}

\newcommand{\gab}{g_{ab}}
\newcommand{\gba}{g_{ba}}
\newcommand{\gae}{g_{ae}}

\newcommand{\pso}{p_{\text{so}}}
\newcommand{\pco}{p_{\text{co}}}

\newcommand{\Pp}{\mathbb{P}}
\newcommand{\vr}{\gamma}

\newtheorem{Proposition}{Proposition}

\newtheorem{Remark}{Remark}

\begin{document}

\title{Artificial Noise Injection for Securing Single-Antenna Systems}


\author{
Biao~He,~\IEEEmembership{Member,~IEEE,}
Yechao~She,~\IEEEmembership{Student Member,~IEEE,}
and
Vincent~K.~N.~Lau,~\IEEEmembership{Fellow,~IEEE}\vspace{-8mm}\\
\thanks{This work was supported by Grant T21-602/15.}
\thanks{B. He is with the Center for Pervasive Communications and Computing, University of California at Irvine, Irvine, CA 92697. Email: biao.he@uci.edu. This work was completed when B. He was with the
Department of Electronic and Computer Engineering, The Hong Kong University of Science and Technology, Hong Kong.}
\thanks{Y. She and V. K. N. Lau are with the Department of Electronic and Computer Engineering, The Hong Kong University of Science and Technology, Hong Kong. Emails:\{yshe, eeknlau\}@ust.hk.}
}

\maketitle

\begin{abstract}
We propose a novel artificial noise (AN) injection scheme for wireless systems over quasi-static fading channels, in which a single-antenna transmitter sends confidential messages to a half-duplex receiver  in the presence of an eavesdropper. Different from  classical AN injection schemes, which rely on a multi-antenna transmitter or external helpers, our proposed scheme is applicable to the scenario where the legitimate transceivers are very simple.
We analyze the performance of the proposed scheme and optimize the design of the transmission. Our results highlight that perfect secrecy is always achievable by properly designing the AN injection scheme.

\end{abstract}

\begin{IEEEkeywords}
Physical layer security, artificial noise, secrecy outage probability.\vspace{-3.5mm}
\end{IEEEkeywords}

\IEEEpeerreviewmaketitle

\section{Introduction}\label{sec:Intro}
The ubiquity of wireless devices in modern life has led to an unprecedented amount of private and sensitive data being transmitted over wireless channels.
Consequently, security issues of wireless transmissions have become critical due to the unalterable open nature of the wireless medium.
As a complement to traditional cryptographic techniques based on  encryption, physical layer security (PLS) has been widely studied for ensuring secure wireless communications by exploiting the characteristics of wireless channels~\cite{Zhou_13_Physical,Yang_15_Mag,Zhao_16_PLSIiIABWNs}.

Secure transmission with artificial noise (AN) injection to confuse eavesdroppers is a key technique for PLS, and has been widely studied in the literature. 
The classical AN injection schemes were investigated in, e.g.,~\cite{Goel_08,Zhou_10,Yang_15_Artif_TC,He_15_Achieving,Yang_16_optimal_TVT}.
For the scenario where the transmitter has multiple antennas, AN is designed to lie  in the nullspace of the receiver's channel  to degrade the eavesdropper's channel, while in the single-antenna scenario, external relays or helpers are adopted to collaboratively generate AN.
In~\cite{Zheng_13_Improving}, an AN injection scheme was proposed for the scenario where the idealized full-duplex receiver is available. The receiver broadcasts AN and receives messages from the transmitter simultaneously.
Although a multi-antenna transmitter and external helpers are not needed, however, the scheme in~\cite{Zheng_13_Improving} requires the advanced full-duplex receiver with very good self-interference cancelation.

Different from the aforementioned scenarios for AN injection schemes, practical wireless systems often consist of simple legitimate transceivers.
A basic system operation is that a single-antenna transmitter wants to send confidential messages to a half-duplex receiver without any helpers.
The simplicity of the legitimate transmitter-receiver pair indeed makes it challenging to inject AN into such a single-antenna system.
For transmission over multi-path fading channels,
AN injection schemes for  single-antenna systems were studied in~\cite{Qin_13_Power,Zhang_16_Artificial,Yang_15_ArtificialN} by exploiting the degrees of freedom from the cyclic prefix. However, these schemes rely on not only the multi-path diversity, but also the idealized assumption that keys or channel state information (CSI) can be secretly shared between the transmitter and the receiver.
To the best of our knowledge, how to effectively inject AN into a basic single-antenna system (over flat fading channels) has still not been addressed in the literature.

In this paper, we propose a novel AN injection scheme for single-antenna systems with a half-duplex receiver and no helpers, which addresses the challenging problem of injecting AN in single-antenna systems. Our proposed scheme does not rely on multi-path diversity or the idealized assumption that keys or CSI are secretly shared. We highlight that perfect secrecy can always be achieved by properly designing the proposed scheme. Note that existing PLS techniques cannot ensure perfect secrecy of transmissions over quasi-static fading channels when the instantaneous CSI of eavesdroppers is unknown. Our results show that the proposed AN injection scheme significantly improves the performance of the single-antenna wiretap system.

Throughout the paper, we adopt the following notations:
$\mathbb{E}\{\cdot\}$ denotes the expectation operation, $\mathbb{P}\{\cdot\}$ denotes the probability measure, $\mathcal{CN}\left(\mu,\sigma^2\right)$ denotes the circularly symmetric complex Gaussian distribution with mean $\mu$ and variance $\sigma^2$.

\vspace{-0mm}
\section{System Model}\label{Sec:SysM}
We consider a wiretap system over quasi-static Rayleigh fading channels. As illustrated in Figure~1, a transmitter, Alice, wants to send confidential messages to a half-duplex receiver, Bob, in the presence of an eavesdropper, Eve. Alice, Bob, and Eve each have a single antenna.\footnote{Although we focus on a single-antenna system in this paper, the proposed AN injection scheme is also applicable to multi-antenna systems where  Alice, Bob, and/or Eve have multiple antennas.}
Note that the consideration of a single-antenna eavesdropper has been widely adopted in the literature, e.g.,~\cite{Zhou_11,He_13_2,Qin_13_Power,He_16_OnMetrics,Hu_16_AVstsithepoo,Hu_17_ANAstcwltfo}.

\begin{figure}[!htb]
\centering
\includegraphics[width=.7\columnwidth]{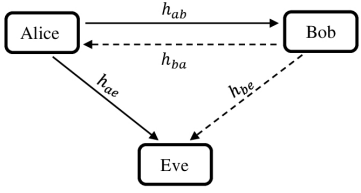}
\vspace{-3mm}
\caption{A wiretap system.}
\vspace{-5mm}  \label{fig:BasicModel}
\end{figure}

\subsection{Channel Model}
We adopt a block fading model where the channel gains remain
constant over a block of symbols and change independently from one
block to the next.
The instantaneous channel gain from $i$ to $j$ ($i,j \in\{a,b,e\}$) is denoted as $h_{ij}\sim\mathcal{CN}\left(0,\mathbb{E}\left\{|h_{ij}|^2\right\}\right)$, where the subscripts $a,b$ and $e$ represent Alice, Bob, and Eve, respectively, and we denote the power gain of the channel from $i$ to $j$ as $g_{ij}=|h_{ij}|^2$. The probability density function (PDF) of $g_{ij}$ is given by
\vspace{-2mm}
\begin{equation}\label{}
  f_{g_{ij}}(g_{ij})=\frac{1}{\bar{g}_{ij}}\exp\left(-\frac{g_{ij}}{\bar{g}_{ij}}\right),
\end{equation}
where $\bar{g}_{ij}=\mathbb{E}\left\{g_{ij}\right\}$ denotes the average power gain of the channel from $i$ to $j$.
We assume that the channel between the transmitter and the receiver is reciprocal, i.e., $h_{ij}=h_{ji}$, and $h_{ab}, h_{ae}$, and $h_{be}$ are independent of each other.
At the start of each block, Alice transmits pilot symbols to enable channel estimation at the receiver.
We assume that Bob can perfectly estimate the channel to Alice, and hence, Bob knows the instantaneous CSI $h_{ab}=h_{ba}$.
We further assume that the duration of a block is sufficiently long and the time spent on training is negligible. 

\vspace{-4mm}
\subsection{Secure Encoding}
We assume that Alice uses the widely adopted wiretap code for data transmissions. There are two rate parameters, namely, the codeword transmission rate, $R_b$, and the confidential information rate, $R_s$. The positive rate difference $R_b-R_s$ is the cost required to provide secrecy against the eavesdropper.
A length $n$ wiretap code is constructed by generating $2^{nR_b}$ codewords $x^n(w,v)$, where $w=1,2,\cdots,2^{nR_s}$ and $v=1,2,\cdots,2^{n(R_b-R_s)}$.
For each message index $w$, we randomly select $v$ from $\left\{1,2,\cdots,2^{n(R_b-R_s)}\right\}$ with uniform probability and transmit the codeword $x^n(w,v)$.
In addition, we consider fixed-rate transmission, where the encoding rates $R_b$ and $R_s$ are fixed over time.

\vspace{-4mm}
\section{New Artificial Noise Injection Scheme}\label{sec:newANsch}
We propose a novel AN injection scheme to defend against eavesdropping attacks in single-antenna systems. The two-phase scheme is detailed as follows.
\vspace{-5mm}
\subsection{Phase 1}
In the first phase, Bob broadcasts pseudo random AN.
The received signal at Alice or Eve in Phase 1 is given by
\vspace{-2mm}
\begin{equation}\label{eq:ya1}
  y_{i, 1}=\sqrt{P_b}\hbi z + n_i, \quad i\in\{a,e\},
\end{equation}
where the subscripts $a$ and $e$ denote the parameters for Alice and Eve, respectively, $z\sim\mathcal{CN}(0,1)$ denotes the normalized complex Gaussian AN from Bob, $P_b$ denotes the average transmit power at Bob, and $n_i\sim\mathcal{CN}(0,\sigma_i^2)$ denotes the additive white Gaussian noise (AWGN) at Alice or Eve.

During Phase~1, Bob does not send any pilot symbols for the channel estimation. Thus, Alice and Eve do not know the instantaneous CSI, i.e., $\hbi$, and cannot decode $z$.

\vspace{-4mm}
\subsection{Phase 2}
In the second phase, Alice forwards the received signal $y_{a, 1}$ from Phase~1 along with the information-bearing signal to Bob.
We denote the normalized transmitted signal at Alice in Phase~2 as $x_a$ with $\mathbb{E}\left\{|x_a|^2\right\}=1$, which is given by
\begin{equation}\label{eq:xa}
  x_a=\sqrt{\alpha} s+\sqrt{1-\alpha}\frac{y_{a, 1}}{\left|y_{a, 1}\right|},
\end{equation}
where $s$ denotes the normalized information-bearing signal, with $\mathbb{E}\left\{|s|^2\right\}=1$, and $0<\alpha\le1$ denotes the power allocation parameter between the information-bearing signal and the AN.
The received signal at Bob or Eve in Phase~2 is given by
\begin{align}\label{eq:yb2}
 &y_{i, 2}=\sqrt{P_a}\hai x_a+n_i=\sqrt{\alpha P_a}\hai s \notag\\
   &+\frac{\sqrt{(1-\alpha) P_a}\hai}{\sqrt{P_b\gba+\sigma_a^2}}\left(\sqrt{P_b}\hba z + n_a\!\right) +n_i,~i\in\{b,e\},  \hspace{-1mm}
\end{align}
where the subscripts $b$ and $e$ denote the parameters for Bob and Eve, respectively, $P_a$ denotes the average transmit power at Alice, and $n_b\sim\mathcal{CN}(0,\sigma_b^2)$ denotes the AWGN at Bob.
From~\eqref{eq:yb2}, we note that Bob needs to know $P_a, P_b, \alpha, \sigma^2_a, \hab=\hba, g_{ba}=|h_{ba}|^2$, and $z$ to cancel the received AN. As mentioned before, Bob knows the instantaneous CSI $\hab=\hba$ and $g_{ba}=|h_{ba}|^2$. He also knows $z$, which is generated by himself in Phase 1, and his average transmit power $P_b$. We further assume that Alice has publicly shared the (fixed) values of $P_a$, $\alpha$, and $\sigma^2_a$ to Bob before the transmission. Thus, Bob can successfully cancel the received AN.
Then, the received signal-to-noise ratios (SNRs) at Bob and Eve are, respectively, given by
\begin{equation}\label{eq:rb}
  \gamma_b=\frac{\alpha P_a\gab}{\frac{(1-\alpha) P_a\gab}{P_b\gba+\sigma_a^2}\sigma_a^2+\sigma_b^2}，
\end{equation}
\begin{equation}\label{eq:re}
  \gamma_e=\frac{\alpha P_a\gae}{(1-\alpha) P_a\gae+\sigma_e^2}.
\end{equation}
It is worth pointing out that \eqref{eq:re} is based on the assumption that Eve has a single antenna. If Eve has multiple antennas, she can take advantage of the receiver diversity to improve the received SNR, while her received SNR is still degraded by the injected AN.

\section{Performance Analysis and Optimal Power Allocation}
\subsection{Performance Analysis}\label{sec:performance analysis_A}
We first derive the PDFs of the received SNRs, and then characterize the security, reliability, and throughput performances of the system. 
\vspace{1mm}
\subsubsection{PDF of Received SNR}
~\\
We rewrite~\eqref{eq:rb} and~\eqref{eq:re} as $\vr_b=\Phi(g_{ab})$ and  $\vr_e=\Psi(g_{ae})$, respectively.
We note that $\Phi(g_{ab})$ and  $\Psi(g_{ae})$ are differentiable and monotonic for $g_{ab}>0$ and $g_{ae}>0$, respectively. Thus, $\vr_b=\Phi(g_{ab})$ and  $\vr_e=\Psi(g_{ae})$ can be uniquely solved for $g_{ab}$ and $g_{ae}$ to give $g_{ab}=\Phi^{-1}(\vr_b)$ and  $g_{ae}=\Psi^{-1}(\vr_e)$, respectively.
We then obtain the PDFs of  $\vr_b$ and $\vr_e$, respectively,~as
\begin{align}\label{eq:pdf_b}
f_{\vr_b}(\vr_b)=&~f_{\gab}\left(\Phi^{-1}\left(\vr_b\right)\right)\frac{\partial\Phi^{-1}\left(\vr_b\right)}{\partial\vr_b} \notag \\
=&~\frac{1}{2\alpha \bar{g}_{ab}P_a P_b}\!\!  \left(\!\frac{\omega_3}{\sqrt{\omega_1^2\!+\!4\alpha\omega_2\vr_b}} \!+\!(1\!-\!\alpha)P_a\sigma_a^2\!+\!P_b\sigma_b^2\!\right)\notag\\
&~\times\exp\left(-\frac{\omega_1+\sqrt{\omega_1^2+4\alpha\omega_2\vr_b}}{2\alpha \bar{g}_{ab}P_a P_b}\right)，
\end{align}
\begin{align}\label{eq:pdf_e}
&f_{\vr_e}(\vr_e)=f_{g_{ae}}\left(\Psi^{-1}\left(\vr_e\right)\right)\frac{\partial\Phi^{-1}\left(\vr_e\right)}{\partial\vr_e}\notag\\
&=\frac{\alpha\sigma_e^2}{P_a\bar{g}_{ae}\left(\alpha\!-\!(1\!-\!\alpha)\vr_e\right)^2}
\exp\!\left(\!-\frac{\vr_e \sigma_e^2}{\bar{g}_{ae} \left(P_a\left(\alpha\!-\!\left(1\!-\!\alpha\right)\vr_e\right)\right)}\right)\!,
\end{align}
where
$\omega_1=\left((1-\alpha)\gamma_b-\alpha\right)P_a\sigma_a^2+P_b\gamma_b\sigma_b^2$,  $\omega_2=P_aP_b\sigma_a^2\sigma_b^2$, and
$\omega_3=\left(\alpha+2(1-\alpha)\vr_b\right)\omega_2+(1-\alpha)\left((1-\alpha)\vr_b-\alpha\right)P_a\sigma_a^4+P_b^2\vr_b\sigma_b^4$.
\vspace{1mm}
\subsubsection{Security, Reliability, and Throughput Performances}
~\\
The security performance is measured by the secrecy outage probability, which is defined by~\cite{Zhou_11}
\begin{equation}\label{}
  \pso=\Pp\left(C_e>R_b-R_s\right),
\end{equation}
where $C_e=\log_2\left(1+\vr_e\right)$ denotes Eve's instantaneous channel capacity.
From~\eqref{eq:pdf_e}, we have
\begin{align}\label{eq:pso1}
\pso
   =&~\left\{\begin{array}{ll}
   \!0\;, &\!\!\!\!\!\!\!\!\!\!\!\!\!\!\!\!\!\!\!\!\!\!\!\!\!\mbox{if}~ \alpha \leq 1-2^{R_s-R_b}, \\ 
   \!\exp\left(-\frac{(2^{R_b}-2^{R_s})\sigma_e^2 }{(2^{R_s}+(\alpha-1)2^{R_b})\bar{g}_{ae} P_a}\right)\;, & \mbox{otherwise.}
   \end{array}\right.
\end{align}

The reliability performance of the system is measured by the connection outage probability, which is defined by
\begin{equation}\label{}
  \pco=\Pp\left(R_b>C_b\right),
\end{equation}
where $C_b=\log_2\left(1+\vr_b\right)$ denotes Bob's instantaneous channel capacity.
From~\eqref{eq:pdf_b}, we have
\vspace{-1mm}
\begin{equation}\label{eq:pco1}
  \pco=\Pp\left(\vr_b<2^{R_b}-1\right)=\int^{2^{R_b}-1}_{0}f_{\vr_b}(\vr_b)~ \mathrm{d} \vr_b.
\end{equation}
The closed-form expression for $\pco$ is intractable due to the complicated expression for $f_{\vr_b}(\vr_b)$ in \eqref{eq:pdf_b}.

The throughput of the system is given by
\begin{equation}\label{eq:eta}
  \eta=\frac{1}{2}\!\left(1\!-\!\pco\right)R_s=\frac{1}{2}\left(1-\int^{2^{R_b}-1}_{0}\!\!f_{\vr_b}(\vr_b) ~\mathrm{d} \vr_b\!\right)\!R_s,
\end{equation}
where the scalar factor $1/2$ is due to the fact that two time units are required in two phases.

\begin{Remark}\label{Remark1}
From~\eqref{eq:pso1}, we highlight that perfect secrecy, i.e., $\pso=0$, can be achieved with the proposed AN injection scheme for any $R_b>R_s>0$ by setting the power allocation parameter as
\begin{equation}\label{}
 \alpha \leq 1-2^{R_s-R_b}.
\end{equation}
Note that existing PLS techniques cannot ensure perfect secrecy of transmissions  over quasi-static fading channels; see, e.g.,~\cite{Goel_08} and~\cite{Zhou_11,He_13_2,He_16_OnMetrics}.

\end{Remark}

\vspace{-2mm}
\subsection{Optimal Power Allocation}\label{sec:optPA}
In the following, we derive the optimal power allocation parameter that maximizes the throughput subject to the security and reliability constraints. We assume that the encoding rates have already been designed.
The problem is formulated as
\begin{subequations}
\begin{eqnarray}\label{eq:designproblem_a}
  \max_{\alpha} &&  \eta\left(\alpha\right) \label{eq:designproblem_obj_a}
  \\
  \text{s.t.}&&   \pso\le\epsilon, \pco\le\delta, 0<\alpha\le1,  \label{eq:designproblem_cons_a}
\end{eqnarray}
\end{subequations}
where $\epsilon\in[0,1]$ and $\delta\in[0,1]$ denote the maximum allowed secrecy outage probability and the maximum allowed connection outage probability, respectively.
We note that $\eta$ monotonically increases as $\alpha$ decreases. Thus, the optimal power allocation parameter is obtained by finding the maximum $\alpha$ that satisfies all the constraints in~\eqref{eq:designproblem_cons_a}, which is given~by
\begin{equation}\label{eq:optimal_alpha_PA}
  \alpha^o=\min\left\{\left(1-2^{R_s-R_b}\right)\left(1-\frac{\sigma_e^2}{ P_a\bar{g}_{ae}\ln\epsilon}\right),1\right\}.
\end{equation}

\section{Joint Rate and Power Allocation Design in Asymptotic Scenario}\label{sec:jointDesign}
In this section, we allow more degrees of freedom, such that $R_b$ and $R_s$ can be optimally chosen, and investigate the joint rate and power allocation design.

\subsection{Problem Formulation}
The design problem is formulated as
\begin{subequations}
\begin{eqnarray}\label{eq:designproblem}
 \!\! \!\! \max_{\alpha, R_b, R_s} &&\eta\left(\alpha,R_b,R_s\right) \label{eq:designproblem_obj}
  \\
 \!\! \!\!  \text{s.t.}&&\!\!\!\!\!\!\!\!\pso\le\epsilon, \pco\le\delta, R_b\ge R_s>0,  0<\alpha\le1.~~  \label{eq:designproblem_cons}
\end{eqnarray}
\end{subequations}
For any given $R_b$ and $R_s$, the optimal $\alpha$ is still given by~\eqref{eq:optimal_alpha_PA}. We find that the closed-form solutions of the optimal $R_b$ and $R_s$ are mathematically intractable due to the complicated expression for the PDF of the received SNR at Bob, i.e., \eqref{eq:pdf_b}. The optimal $R_b$ and $R_s$ for the design problem can be obtained only by numerically solving  $\displaystyle{\max_{R_b, R_s} \eta\left(\alpha=\alpha^o,R_b,R_s\right)}$ subject to the constraints in \eqref{eq:designproblem_cons}.
In the following subsections, we study an asymptotic scenario where  $\sigma_a^2\rightarrow0$, in which the closed-form solutions of $R_b$ and $R_s$ are tractable.
\subsection{Asymptotic Analysis}
We now resort to the asymptotic analysis of the scenario of $\sigma_a^2\rightarrow0$. 
This condition can be validated when Alice has a very sensitive receiver compared with Bob, such that $\sigma_a^2\ll\sigma_b^2$.
In such a scenario, the noise power at Bob is determined by $\sigma_b^2$ only,
and the received SNR at Bob is rewritten as
\begin{equation}\label{eq:rb_sc}
  \gamma_b=\alpha P_a\gab/\sigma_b^2.
\end{equation}
Then, the PDF of $\vr_b$ becomes
\begin{equation}\label{}
  f_{\vr_b}(\vr_b)=\frac{\sigma_b^2}{\alpha P_a \bar{g}_{ab}}\exp\left(-\frac{\sigma_b^2\vr_b}{\alpha P_a\bar{g}_{ab}}\right),
\end{equation}
the connection outage probability becomes
\begin{equation}\label{eq:pco2}
  \pco=\Pp\left(\vr_b<2^{R_b}-1\right)=1-\exp\left(-\frac{\sigma_b^2\left(2^{R_b}-1\right)}  {\alpha P_a\bar{g}_{ab} } \right),
\end{equation}
and the throughput of the system becomes
\begin{equation}\label{eq:eta_2}
  \eta=\frac{1}{2}\left(1-\pco\right)R_s=\frac{1}{2}\exp\left(-\frac{\sigma_b^2\left(2^{R_b}-1\right)}  {\alpha P_a\bar{g}_{ab} }\right)R_s.
\end{equation}
The expressions for $\vr_e$, $f_{\vr_e}(\vr_e)$, and $\pso$ are kept unchanged as \eqref{eq:re}, \eqref{eq:pdf_e}, and \eqref{eq:pso1}, respectively.

\subsection{Feasible Constraint}
We first determine the feasible constraints subject to which a non-zero throughput is achievable. The feasible range of $\epsilon$ and $\delta$ in the asymptotic scenario where $\sigma_a^2\rightarrow0$ is summarized in the following proposition.
\begin{Proposition}\label{pro:feasible}
The feasible range of the security and reliability constraints is given by
\begin{equation}\label{eq:feasibleRanges}
  \left\{(\epsilon,\delta) : 0\le\epsilon\le1, \delta_l(\epsilon)<\delta\le1 \right\},
\end{equation}
where
\begin{align}\label{eq:delta_lower_B}
\delta_l(\epsilon)=1-\exp\left(-\frac{\sigma_b^2}{P_a\bar{g}_{ab}\left(1-\frac{\sigma_e^2}{P_a\bar{g}_{ae}\ln\epsilon}\right)}\right).
\end{align}
\end{Proposition}
\begin{IEEEproof}
See Appendix~\ref{AP:feasible}.
\end{IEEEproof}

\subsection{Design Solution}
The closed-form solutions of the optimal $\alpha$, $R_b$, and $R_s$ to the design problem in the asymptotic scenario where $\sigma_a^2\rightarrow0$ are summarized in the following proposition.

\begin{Proposition}\label{Prop:optimalarbs}
The optimal $\alpha$, $R_b$, and $R_s$ that maximize the throughput subject to the security and reliability constraints  are given by 
\begin{equation}\label{eq:opt_a_prop2}
  \alpha^*=\left(1-2^{R_s^*-R_b^*}\right)\left(1-\frac{\sigma_e^2}{ P_a\bar{g}_{ae}\ln\epsilon}\right),
\end{equation}
\begin{equation}\label{eq:opt_Rb}
  R_b^*=\min\left\{\log_2\left(\phi_1\right), \log_2\left(2^\psi+\sqrt{4^\psi-2^\psi}\right)\right\},
\end{equation}
\begin{equation}\label{eq:opt_Rs}
  R_s^*=\min\left\{\log_2(\phi_2), \psi\right\},
\end{equation}
where
\begin{equation}\label{}
  \phi_1=\frac{ \sigma_b^2\bar{g}_{ae}\ln\epsilon+\bar{g}_{ab}\ln\left(1-\delta\right)\left(\sigma_e^2- P_a\bar{g}_{ae}\ln\epsilon\right)}{2\sigma_b^2\bar{g}_{ae}\ln\epsilon},
\end{equation}
\begin{equation}\label{}
  \phi_2=-\frac{\left(\sigma_b^2\bar{g}_{ae}\ln\epsilon\!+\!\bar{g}_{ab}\ln\left(1\!-\!\delta\right)\left(\sigma_e^2- P_a\bar{g}_{ae}\ln\epsilon\right)\right)^2}{4\sigma_b^2\bar{g}_{ab}\bar{g}_{ae}\ln\left(1-\delta\right)\ln\epsilon\left(P_a\bar{g}_{ae}\ln\epsilon-\sigma_e^2\right)},
\end{equation}
and $\psi$ is the solution of $x$ to
\begin{equation}\label{}
  \frac{\sqrt{4^x-2^x}}{x2^x\left(2^{x+1}\!+\!2\sqrt{4^x-2^x}\!-\!1\right)}=\frac{\sigma_b^2\bar{g}_{ae}\ln\!2\ln\epsilon}{\bar{g}_{ab}\left(P_a\bar{g}_{ae}\ln{\epsilon}-\sigma_e^2\right)}.
\end{equation}
\end{Proposition}
\begin{IEEEproof}
See Appendix~\ref{AP:optimalsolutions}.
\end{IEEEproof}

\section{Numerical Results}\label{sec:numerres}

\begin{figure}[!t]
\centering
\includegraphics[height=1.6in,width=1.9in]{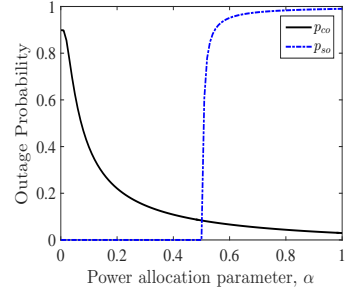}
\caption{The outage provability versus the power allocation parameter. The system parameters are $R_b=2, R_s=1,$ $P_a=P_b=10$ dB, $\bar{g}_{ab}=\bar{g}_{ae}=1$, and $\sigma_a^2=\sigma_b^2=\sigma_e^2=0.1$.}
\vspace{0mm}  \label{fig:poutvsalpha}
\end{figure}

We first show the reliability and security performances of the proposed AN injection scheme with different power allocations. Figure~\ref{fig:poutvsalpha} plots the connection outage probability, $\pco$, and the secrecy outage probability, $\pso$, versus the power allocation parameter, $\alpha$. The encoding rates are fixed at $R_b\!=\!2$ and $R_s\!=\!1$. As depicted in the figure, $\pco$ decreases as $\alpha$ increases, while $\pso$ increases as $\alpha$ increases. This observation indicates that there exists a tradeoff between the reliability and security performances. Allocating more power to the information-bearing signal and less power to the AN improves the reliability performance but worsens the security performance. Additionally, we find that $\pso\!=\!0$ when $\alpha\!\le\!0.5$ in the figure, which confirms that perfect secrecy is achievable by the proposed AN injection scheme with enough power allocated to the AN. Note that $\pso\simeq1$ when $\alpha=1$, which indicates that the system is very insecure without the proposed AN injection scheme.

\begin{figure}[t!]
    \centering
    \begin{subfigure}[t]{0.5\columnwidth}
        \centering
        \includegraphics[height=1.57in,width=1.85in]{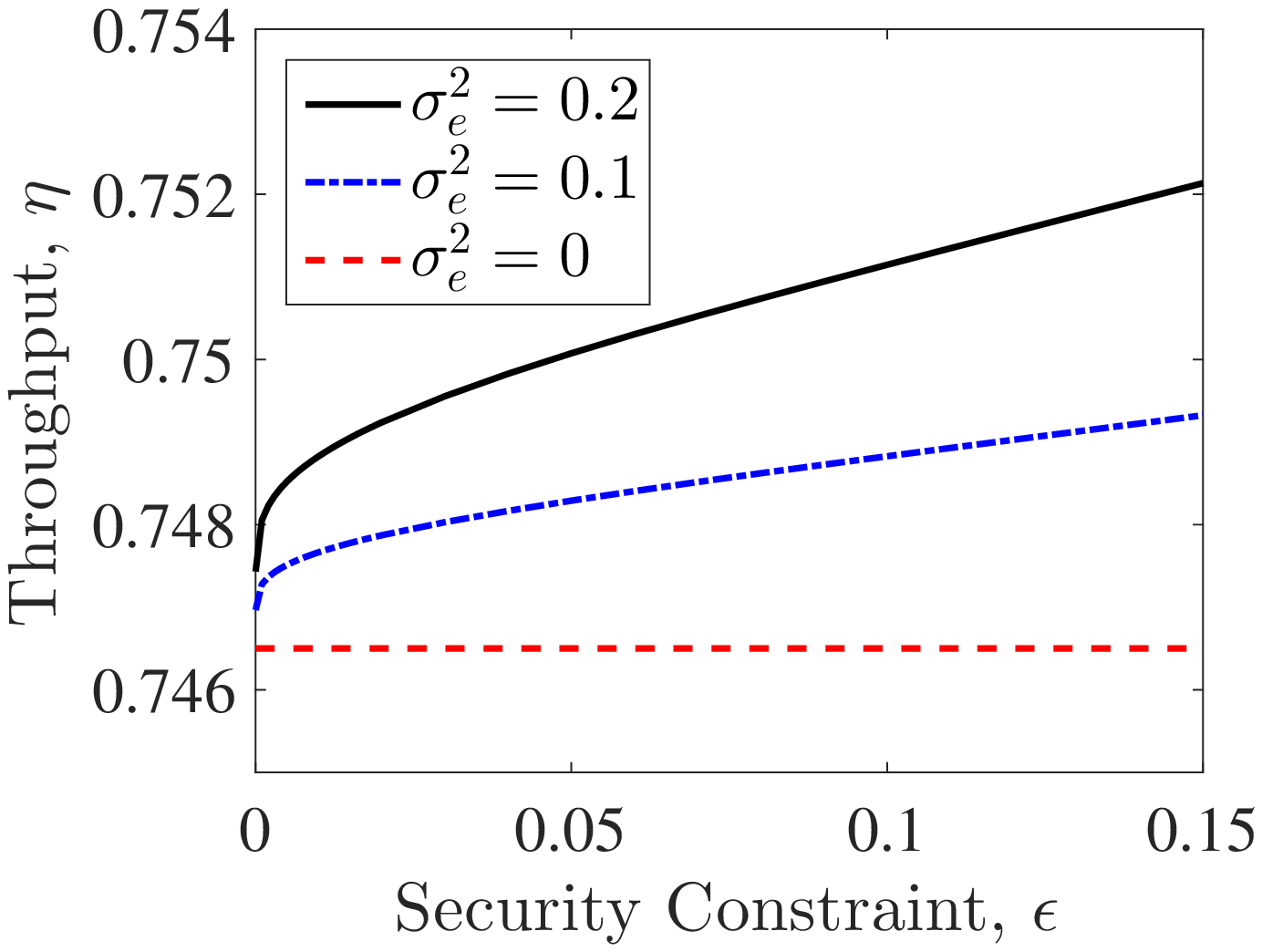}
        \caption{Proposed scheme.} \label{fig:EtaVsEps_AN}\vspace{-1mm}
    \end{subfigure}%
    ~\begin{subfigure}[t]{0.5\columnwidth}
        \centering
        \includegraphics[height=1.57in,width=1.85in]{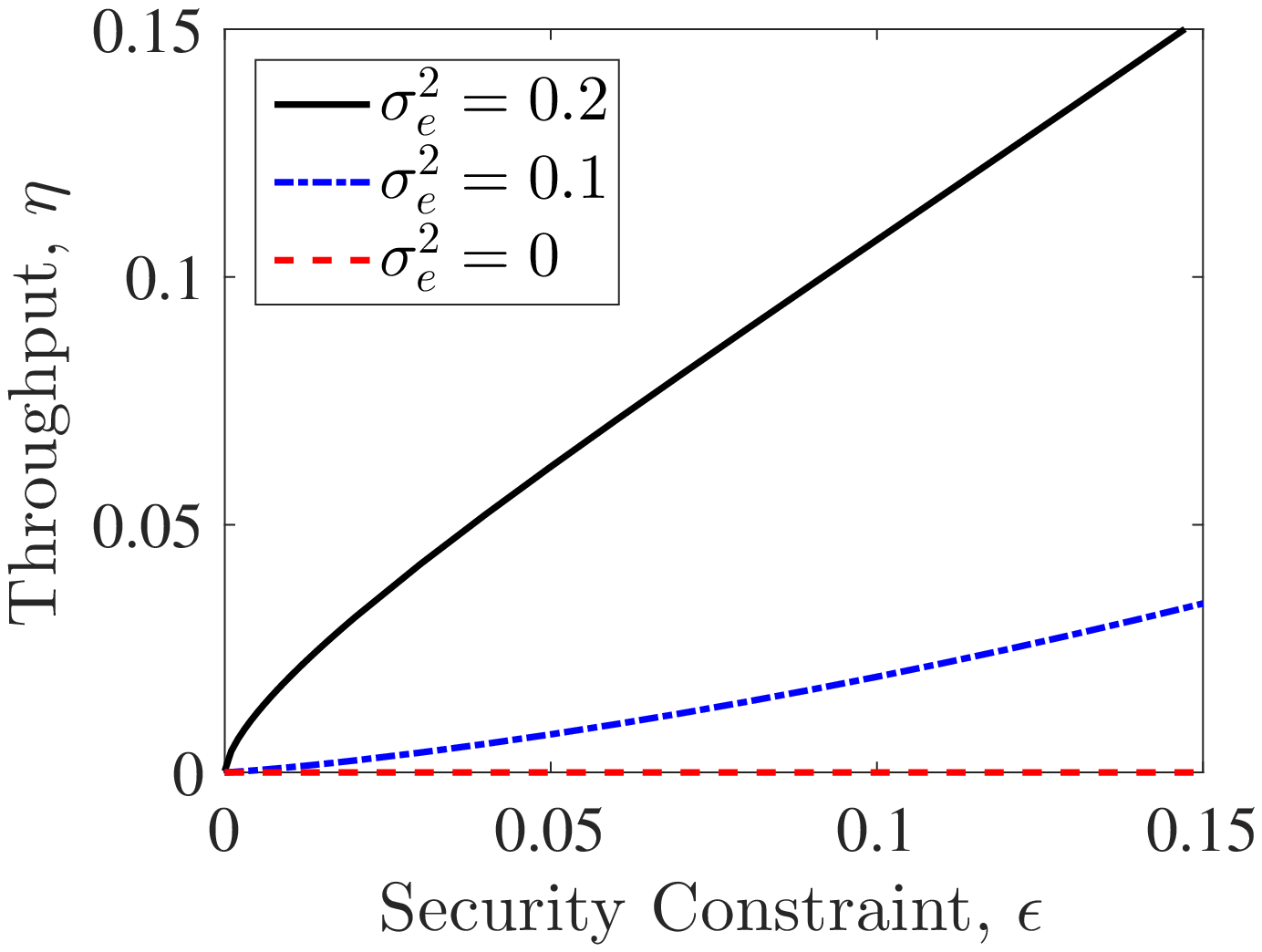}
        \caption{Benchmark scheme.} \label{fig:EtaVsEps_BM}\vspace{-1mm}
    \end{subfigure}
    \caption{The throughput versus the security constraint. The system parameters are  $P_a=P_b=10$ dB, $\bar{g}_{ab}=\bar{g}_{ae}=1, \sigma_a^2=0, \sigma_b^2=0.1, \sigma_e^2=0, 0.1, 0.2$, and $\delta=0.1$.} \label{fig:etavseps} \vspace{-5mm}
\end{figure}

We now compare the performances of our proposed AN injection scheme and a benchmark scheme.
We adopt the secure on-off transmission scheme~\cite{Zhou_11,He_13_2}, which is an existing secure transmission scheme for single-antenna systems, as the benchmark scheme. The comparison results are  presented in Figure~\ref{fig:etavseps}, which plots the throughput, $\eta$, versus the security constraint, $\epsilon$. The performance achieved by the proposed AN injection scheme is shown in Figure~\ref{fig:etavseps}(a), and the performance achieved by the benchmark scheme is shown in Figure~\ref{fig:etavseps}(b). We note that our proposed scheme always significantly outperforms the benchmark scheme. In particular, the case of $\sigma_e^2=0$ represents the scenario where Eve has a very sensitive receiver, which implies the worst-case consideration (from the legitimate users' point of view) when $\sigma_e^2$ is unknown, and the secrecy constraint of $\epsilon=0$ indicates the requirement of perfect secrecy. We highlight that non-zero throughput is achievable by the proposed AN injection scheme even with the worst-case consideration $\sigma_e^2=0$ and the requirement of perfect secrecy $\epsilon=0$. In contrast, non-zero throughput is not achievable by the existing benchmark scheme with either   the worst-case consideration or the requirement of perfect secrecy.

\balance

\section{Conclusion and Future Work}\label{sec:con}
In this paper, we have proposed a novel AN injection scheme which is applicable to a single-antenna system without any helpers. We have analyzed the scheme performance and  optimized the power allocation. Furthermore, we have investigated the joint rate and power allocation design in the asymptotic scenario where $\sigma_a^2\rightarrow0$. Our results show that the proposed AN injection scheme effectively improves the performance of the system, and even perfect secrecy is achievable by allocating enough power to the AN. It is worth mentioning that the application of the proposed AN injection scheme is not limited to single-antenna systems, and an interesting future research direction is to evaluate the performance of the proposed scheme in multi-antenna systems. 

\appendices
\section{Proof of Proposition~\ref{pro:feasible}}\label{AP:feasible}
The feasible security constraint is $0\le\epsilon\le1$ without the consideration of the reliability constraint.
The problem to find the minimum achievable $\pco$ is formulated as
\begin{subequations}
\begin{eqnarray}\label{}
  \min_{\alpha, R_b, R_s} &&  \pco   \\
  \text{s.t.}&&  \pso\le\epsilon, R_b>R_s>0, 0<\alpha\le1.
\end{eqnarray}
\end{subequations}
From \eqref{eq:pco2}, we find that $\pco$ is a decreasing function of $\alpha$ for any given $R_b$. Then, for any given $R_b$ and $R_s$, it is wise to have the maximum $\alpha$ that satisfies $\pso\le\epsilon$ and $0<\alpha\le1$, which is given in~\eqref{eq:optimal_alpha_PA}.
For $R_b$ and $R_s$ satisfying $\left(1-2^{R_s-R_b}\right)\left(1-\sigma_e^2/\left(\bar{g}_{ae} P_a\ln\epsilon\right)\right)\leq 1$, it is wise to choose
\begin{equation}\label{eq:alpha_conc1}
  \alpha=(1-2^{R_s-R_b})(1-\frac{\sigma_e^2}{\bar{g}_{ae} P_a\ln\epsilon}).
\end{equation}
Substituting \eqref{eq:alpha_conc1} into \eqref{eq:pco2}, we find that $\pco(R_b,R_s)$ is a convex function with respect to (w.r.t.) $R_b$ for any given $R_s$, and it is wise to have
\begin{equation}\label{eq:Rb_con}
R_b=\log_2(2^{R_s}+\sqrt{4^{R_s}-2^{R_s}})
\end{equation}
to minimize $\pco$.
Substituting \eqref{eq:alpha_conc1} and \eqref{eq:Rb_con} into \eqref{eq:pco2},
we find that  $\pco(R_s)$ is a decreasing function of $R_s$. We then find that the minimum achievable $\pco$ for $R_b$ and $R_s$ that satisfy $\left(1-2^{R_s-R_b}\right)\left(1-\sigma_e^2/\left(\bar{g}_{ae} P_a\ln\epsilon\right)\right)\leq 1$ approaches \eqref{eq:delta_lower_B} by calculating $\lim_{R_s\rightarrow0}\pco(R_s)$.
Following similar steps to those described above, we find that the minimum achievable $\pco$ for $R_b$ and $R_s$ that satisfy $\left(1-2^{R_s-R_b}\right)\left(1-\sigma_e^2/\left(\bar{g}_{ae} P_a\ln\epsilon\right)\right)> 1$ is always larger than~\eqref{eq:delta_lower_B}. Thus, the minimum achievable $\pco$ approaches~\eqref{eq:delta_lower_B} for $R_b\!\ge\! R_s\!>\!0$. This completes the proof.

\section{Proof of Proposition~\ref{Prop:optimalarbs}}\label{AP:optimalsolutions}
For any given $R_b$ and $R_s$, the optimal $\alpha$ is given by~\eqref{eq:optimal_alpha_PA}. For $R_b$ and $R_s$ that satisfy $\left(1-2^{R_s-R_b}\right)\left(1-\sigma_e^2/\left(\bar{g}_{ae} P_a\ln\epsilon\right)\right)\leq 1$, it is wise to have $\alpha$ as \eqref{eq:alpha_conc1}.
Substituting \eqref{eq:alpha_conc1} into \eqref{eq:eta_2}, we find that $\eta\left(R_b,R_s\right)$ is a concave function w.r.t. $R_b$ for any given $R_s$, and it is wise to have $R_b$ as \eqref{eq:Rb_con} to maximize $\eta$.
Then, substituting \eqref{eq:alpha_conc1} and \eqref{eq:Rb_con} into \eqref{eq:eta_2},
we find that  $\eta\left(R_s\right)$ is a concave function w.r.t. $R_s$ for $R_s>0$. With the consideration of $R_s\le R_b$, we obtain the optimal $R_s$ as \eqref{eq:opt_Rs} and the corresponding optimal $R_b$ as \eqref{eq:opt_Rb}.
Following similar steps to those described above, we find that the maximum achievable $\eta$ for $R_b$ and $R_s$ that satisfy $\left(1-2^{R_s-R_b}\right)\left(1-\sigma_e^2/\left(\bar{g}_{ae} P_a\ln\epsilon\right)\right)> 1$ is always smaller than the $\eta$ achieved by having $R_b$ and $R_s$ as \eqref{eq:opt_Rb} and \eqref{eq:opt_Rs}. Thus, the solutions of the optimal $\alpha$, $R_b$, and $R_s$ are given as \eqref{eq:opt_a_prop2}, \eqref{eq:opt_Rb}, and~\eqref{eq:opt_Rs}, respectively. 
This completes the proof.


\end{document}